\documentclass[letterpaper,12pt]{article}

\usepackage{dm}

\begin{document}

\makeatletter
\renewcommand{\fnum@figure}{\scriptsize\textbf{\figurename~\thefigure}}
\renewcommand{\refname}{}
\makeatother

\vspace*{1\baselineskip}

\begin{center}
{\bf \Large \scshape 
Dark Matter Structures in the Universe: \\
\medskip
\large Prospects for Optical Astronomy in the Next Decade\\
\medskip
{\it \footnotesize Submitted to the 2010 Astronomy \& 
Astrophysics Decadal Survey panel} }
\end{center}

\vspace*{1\baselineskip}

\begin{center}
{\bf 
P.~J.~Marshall,\footnote{\scriptsize Physics Department, University of California, Santa Barbara}$^{,}$\footnote{\scriptsize Primary contact: \texttt{pjm@physics.ucsb.edu, +1 805 893 3189}}
M.~Auger,$^1$ 
J.~G.~Bartlett,\footnote{\scriptsize APC, Universit\'e Paris Diderot, France} 
M.~Brada\v{c},$^1$ 
A.~Cooray,\footnote{\scriptsize Department of Physics, University of California, Irvine} 
N.~Dalal,\footnote{\scriptsize CITA, University of Toronto}\\
G.~Dobler,\footnote{\scriptsize Institute for Theory and Computation, Harvard University}
C.~D.~Fassnacht,\footnote{\scriptsize Department of Physics, University of California, Davis} 
B.~Jain,\footnote{\scriptsize Department of Physics \& Astronomy, University of Pennsylvania} 
C.~R.~Keeton,\footnote{\scriptsize Department of Physics \& Astronomy, Rutgers University} 
R.~Mandelbaum,\footnote{\scriptsize Institute for Advanced Study, Princeton}
L.~A.~Moustakas,\footnote{\scriptsize Jet Propulsion Laboratory, Pasadena}
M.~A.~Strauss,$^{10}$
J.~A.~Tyson,$^7$ 
D.~Wittman,$^7$ 
S.~A.~Wright$^4$ 
}
\end{center}


\vspace*{1\baselineskip}

\begin{center}
\begin{minipage}{0.85\linewidth}
{\it \footnotesize
The Cold Dark Matter theory of gravitationally-driven hierarchical
structure formation has earned its status as a paradigm by explaining
the distribution of matter over large spans of cosmic distance and
time.  However, its central tenet, that most of the matter in the
universe is dark and exotic, is still unproven; the dark matter
hypothesis is sufficiently audacious as to continue to 
warrant a diverse battery of tests.  
While local searches for dark matter particles or
their annihilation signals could prove the existence of the substance
itself, studies of cosmological dark matter {\upshape in situ} are 
vital to fully
understand its role in structure formation and evolution.  We argue that
gravitational lensing provides the cleanest and farthest-reaching
probe of dark matter in the universe, which can be combined with other
observational techniques to answer the most challenging and exciting
questions that will drive the subject in the next decade: What is the
distribution of mass on sub-galactic scales?  How do galaxy disks form
and bulges grow in dark matter halos?  How accurate are CDM
predictions of halo structure?  Can we distinguish between a need for
a new substance (dark matter) and a need for new physics (departures
from General Relativity)?  What is the dark matter made of anyway?  We
propose that the central tool in this program should be a wide-field
optical imaging survey, whose true value is realized with support in the
form of high-resolution, cadenced optical/infra-red imaging, and
massive-throughput optical spectroscopy.
}
\end{minipage}
\end{center}

\vspace*{1\baselineskip}

\thispagestyle{empty}

\newpage
\setcounter{page}{1}


\subsection*{Introduction}

It is remarkable that the conceptually-simplest numerical model of structure
formation in the universe -- an N-body simulation, where each particle
interacts with others only via gravity -- provides such a good
match to the distributions of galaxies, groups and clusters we observe
\citep[\eg][]{Teg++04}. The fact that the initial conditions for this  Cold
Dark Matter  model can be set by a handful of parameters constrained 
self-consistently by multiple cosmological datasets \citep[\eg][]{ACBAR,WMAP5}
fully justifies its status as the ``standard model'' of cosmology.

\begin{figure*}[!t]
\centering
\begin{minipage}{0.55\linewidth}
\centerline{\includegraphics[width=0.9\linewidth]{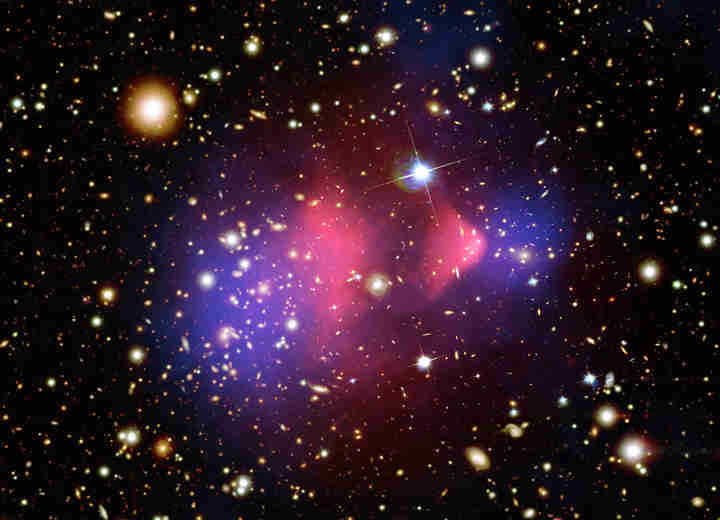}}
\end{minipage}\hfill
\begin{minipage}{0.4\linewidth}
\centerline{\includegraphics[width=0.9\linewidth]{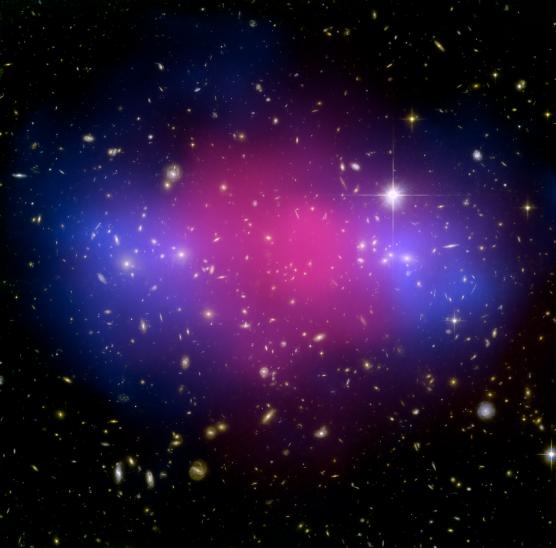}}
\end{minipage}
\caption{The color composite of the Bullet cluster \protect{\bulletcluster}
 (left) and  {\babybullet} (right). Overlaid in {\it blue} shade is the surface mas
s density map from the weak lensing mass reconstruction. 
The X-ray emitting plasma is shown in {\it red}. Both images subtend $\sim 10$~arcmin
on the vertical axis.
Credit 
(left): X-ray NASA/CXC/CfA Optical: NASA/STScI;
  Magellan/U.Arizona; 
  \protect{\citet{Clo++06,Bra++06}} 
(right) X-ray (NASA/CXC/Stanford/S.Allen); 
  Optical/Lensing (NASA/STScI/UCSB/M.Bradac);  
  \protect{\citet{Bra++08}}.}
\label{fig:bullet}
\end{figure*}

The unsettling aspect of this standard model is that $\sim80$\% of the matter 
in it is of unknown form.  This fact has led a number of
researchers to look for alternative explanations for the phenomena attributed
to dark matter: if we instead claim to understand the stress-energy tensor
right-hand side of the Einstein equation of General Relativity, so the
argument goes, then the left-hand side must be adjusted. However, such
``modified gravity'' theories that do not include dark matter have yet to be
shown to be able to fit all  the data.  A striking example of this is provided
by the ``bullet''  clusters studied by \citet{Clo++06} and 
\citeauthor{Bra++06}~(\citeyear{Bra++06},\citeyear{Bra++08}). In these 
high-speed,  plane-of-sky collisions, the positions
of the gravitational potential wells -- mapped using both the observed strong
and weak gravitational lensing effects --  and the dominant baryonic (stellar
and hot gas) mass distributions, are well separated, leading us to infer the
clear presence and domination of a dark matter component (see
Figure~\ref{fig:bullet}).  Note the key role played by gravitational lensing
in this case study: it is the cleanest probe of mass
distributions in the universe, allowing them to be mapped and modeled (albeit
in projection) while avoiding the confusion introduced
by tracers which may or may not be in equilibrium in their potential wells. 

While it is hard to reconcile these lensing results with simply-modified gravity
and no dark matter, this does not mean that GR holds in other regimes.  
Testing the laws of gravity on the largest scales,
where the universal expansion appears to be accelerating, will be a major
activity in the next decade, but many of these tests will rely on the CDM
framework for understanding large scale structure. {\it It is not possible to
fully test gravity theories, or probe dark energy, without understanding the  
behavior of dark matter}.

On the largest scales then, CDM has been initially successful. 
What are the outstanding problems that can be approached using
astronomical observations? 
In this paper, we suggest the following
four questions:
\begin{itemize}
  \item What is the distribution of mass on sub-galactic scales?  
  \item How do galaxy disks form and bulges grow in dark matter halos?  
  \item How accurate are CDM predictions of halo structure?  
  \item Can we distinguish between a need for a new substance (dark matter) and a 
  need for new physics (departures from General Relativity)?  
\end{itemize}

\noindent In each of these areas we can look for clues relating to  the single
most important question about dark matter: {\it what is it?}  Direct searches
for dark matter are underway: some use solid state devices deep underground to
detect the dark matter wind resulting from our motion through our own galaxy's
halo.  Others involve looking for DM particle annihilation signals, coming
from nearby galaxy cores predicted to host the required densities, in the
gamma-ray part of the EM spectrum. 
However, there is a need to extrapolate the results of these experiments to
the universe outside the local group:  we seek a complete understanding of
dark matter, connecting the largest scales to the smallest, over the whole
span of cosmic time. 
Distinguishing between different proposed types of dark matter
(self-interacting, decaying, \etc \etc) will take experiments on many
different length and mass scales, and so requires the study of more halos that
just the one that we live in.
We recommend here a program of gravitational  investigations into structures
at cosmological distance as a necessary complement to the local studies.

This paper is structured as follows.  We focus on three approximate length
scales, and explore in more detail a battery of dark matter experiments  we
would like to carry out at each one, based on the questions listed above.  We
sketch out the facilities required to carry out each experiment, and finally
summarize our findings.


\subsection*{Clusters and groups of galaxies}

\begin{figure*}[!t]
\centering
\begin{minipage}{0.35\linewidth}
\centerline{\includegraphics[width=\linewidth]{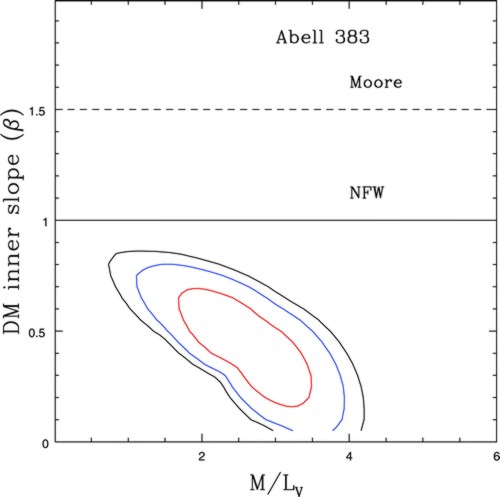}}
\end{minipage}\hfill
\begin{minipage}{0.33\linewidth}
\centerline{\includegraphics[width=\linewidth]{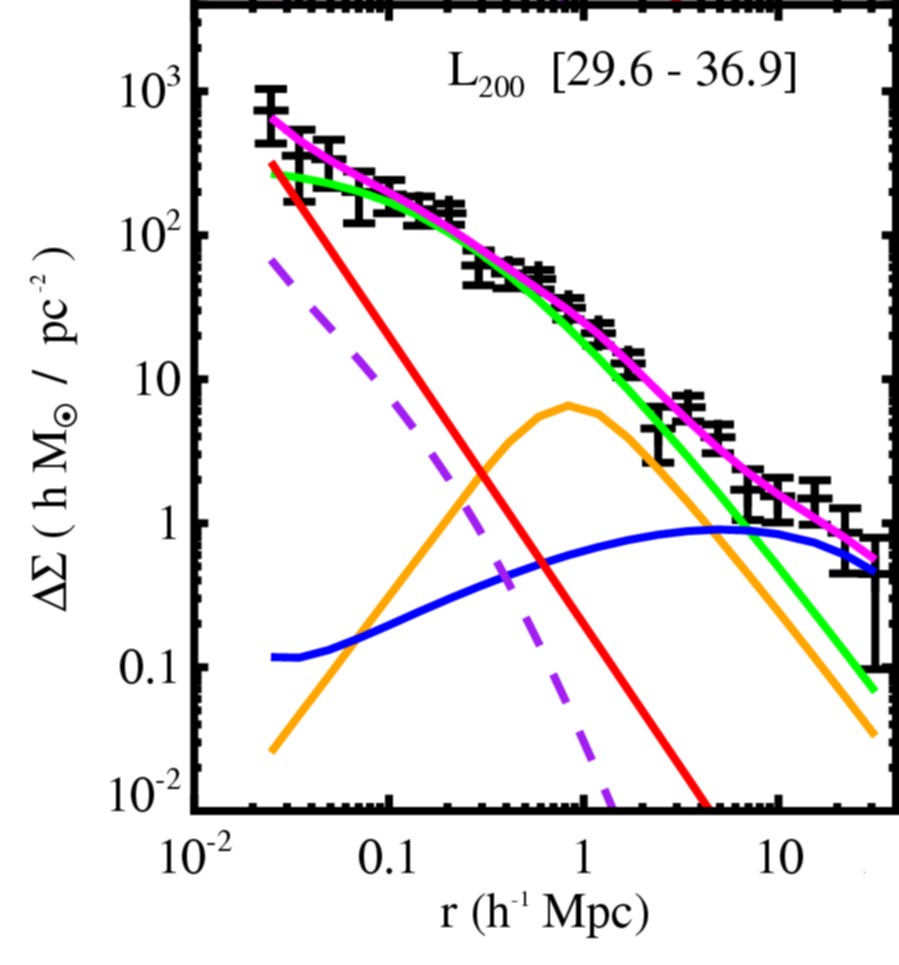}}
\end{minipage}\hfill
\begin{minipage}{0.29\linewidth}
\centerline{\includegraphics[width=\linewidth]{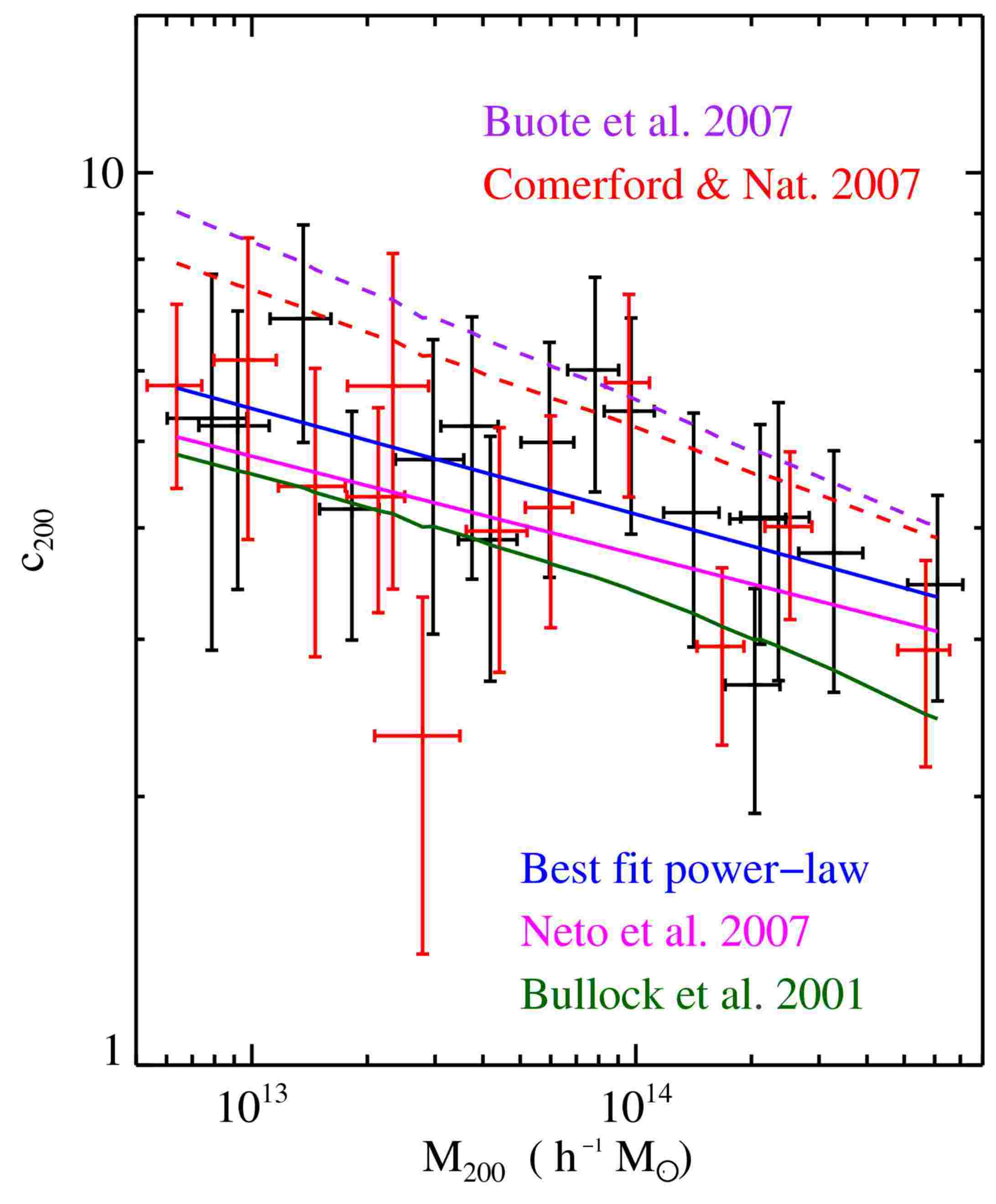}}
\end{minipage}
\caption{Recent results on cluster structure. Left: measuring the inner
density profile  slope in a detailed analysis of a strong lensing cluster, 
including the stellar kinematics in the BCG \citep{San++08}. Center: stacking
optically-selected cluster shear signals measured in the SDSS imaging
\citep{Joh++07} allows the concentration-mass relation to be probed (right).
}
\label{fig:clusters}
\end{figure*}

On cluster scales, the key questions are: How accurate are CDM predictions of
halo structure? and Can we distinguish between dark matter and departures from
General Relativity? One of the most accessible predictions of CDM N-body
simulations has been, and will continue to be as simulations improve,  that of
a universal density profile of the collapsed halos  \citep[\eg ][]{NFW97}, and
how its parameters change with mass scale and time \citep[\eg][]{Bul++01}.
Clusters of galaxies are good places to try and measure this, since they are
dark matter-dominated systems whose size and mass lead to relatively easily
measured gravitational lensing effects. However, to date only a few dozen
clusters have been measured in detail \citep[see  \eg][for an example
compilation]{C+N07}. 
Initial results are intriguing, with massive clusters seeming to be more
concentrated \citep[\eg ][]{Ume++05}, or to have shallower inner slopes
\citep[][Figure~\ref{fig:clusters}]{San++08}, than predicted.  This test of
CDM is still in its infancy, and more work (both observational and
theoretical) is needed to disentangle the effects of baryons on the dark
matter profiles.  
Large numbers of well-measured clusters are needed: not only to ensure a fair
selection (the easiest clusters to measure may not be representative of the
population), but also to  accurately probe the {\it distribution} of density
profiles, and to allow us to compare their {\it evolution} to that predicted
by the simulations.  

A complementary route towards understanding CDM halo profiles is to stack the
weak lensing signals of large numbers of clusters, a technique shown to be
very powerful even using the low-resolution SDSS imaging \citep[in, \eg the
MaxBCG project,][]{Joh++07}. Smaller mass scales are accessible with this
approach, which also allows the mean density profile at very large radii to be
measured and compared against CDM predictions of $\rho \sim r^{-3}$ outside
the virial radius.  
Moreover, a comparison of the stacked weak lensing signal to the  stacked
velocity dispersion profile in the same cluster sample will also test  gravity
on cluster scales.  Dynamics (velocities) are sensitive to one  metric
potential $\Psi$, while lensing measures the sum of the metric  potentials
$\Psi+\Phi$.  In General Relativity, $\Phi=\Psi$, but this is  not the case
for many modified gravity theories.  \cite{BRB06} performed  an analogous test
of gravity on kpc scales using a handful of strong lens galaxies; a much
larger cluster sample would extend the bounds up to  Mpc scales and beyond.  A
key requirement for this will be an intensive  spectroscopic follow-up program
to measure the velocities of cluster members.

Plane-of-sky merging clusters (Figure~\ref{fig:bullet}) can be used to  help
answer the question of what the dark matter is,  placing constraints on the
self-interaction cross-section of the putative dark matter particle itself:
currently $\sigma/m \lesssim 0.7\mbox{cm}^2\mbox{g}^{-1}$ \citep{Ran++08}. The
most stringent constraints come from comparing the relative positions of the
centers of  total mass (from gravitational lensing) and stellar mass (from
optical imaging), indicating  that we can consider expanding this study to use
many more clusters as dark matter laboratories in this way.   Only a fraction
of clusters will be suitable: large samples of objects are needed to find the
most powerful examples.  Systematic errors dominate: in this, and the profile
measurement test, closer comparisons with more physically-realistic
simulations will be needed.

A $\sim10^4$~square degree optical imaging survey, such as that envisioned
with JDEM or LSST, will naturally contain a sample of clusters large enough to
carry out the tests described here with close-to-ultimate precision, provided
it is deep enough (and red enough) to allow accurate measurement of the weak
lensing  signal behind clusters at $z \simeq 1$ (a challenge for the
photometric redshift calibration). It is the strong and weak lensing
combination that is key in constraining the cluster density profile
\citep[\eg][]{Bra++08b}; strong lensing demands high ($\sim 0.1$") resolution
multi-filter imaging, needed to identify candidate multiple-image systems (HST
really enabled the beginning of this work in the last decade). The next
generation AO facilities will provide a viable alternative to HST and JWST in
providing the high resolution images needed. Measuring redshifts for the
multiple-image systems is also vital in pinning down the density profiles in
the core -- as is accounting for the stellar mass in the BCG via its stellar 
kinematics. With most source emission lines and cluster galaxy absorption
lines falling in the optical, high-throughput multi-object spectroscopy with
8-10m class telescopes will continue to be important in building significant
samples of density-profile clusters.


\subsection*{Galaxies}

Galaxy-scale dark matter studies are  complicated by the greater role played
by baryonic matter -- stars and gas -- in the structure and evolution of their
halos. This is an opportunity though -- we would very much like to understand
how galaxies form and develop! CDM provides a robust framework for
understanding galaxies: improved observations are driving an iterative cycle
of ever-improving hydrodynamical simulations and  semi-analytic modeling,
enabling us to learn more about the astrophysical processes going on. 
However, as with clusters, gravitational lensing studies of galaxies can
provide important insights into the dark matter properties of galaxies,
providing accurate measurements of total projected mass, independent of their
luminous properties or dynamical state. The key questions are: How accurate
are CDM predictions of galaxy halo structure? and  How do galaxy disks form
and bulges grow in dark matter halos?

Weak gravitational lensing provides information on the average  halo density
profile at large galacto-centric radii \citep[\eg
][]{Hoe++05,Man++06a,Hey++06}.   Strong lensing, which probes smaller radii,
provides an excellent complement to weak lensing investigations of galaxy mass
distributions \citep[\eg][]{Gav++07}. Further mass measurements, such as those
available from stellar dynamics \citep[\eg][]{T+K04}, lens image time delays
(in lens systems where the source is time-variable), and microlensing
densitometry \citep[if the strongly-lensed source is a quasar,
\eg][]{Poo++08,Mor++08}, can provide valuable additional constraints. Even
stronger constraints can come from compound lens systems in which two (or
more) background objects are strongly lensed by the same foreground galaxy
\citep[\eg, SDSSJ0946$+$1006][]{Gav++08} -- the multiple source planes provide
two high-precision mass estimates at different radii: one out of every 40-80 
elliptical lenses will have multiple sources detectable in
deep, high resolution imaging.

The above approaches have already produced important results.  For example, 
massive elliptical galaxies have been found to be a remarkably structurally
homogeneous population, with isothermal ($\rho(r) \propto r^{-2}$) total
density profiles over a range of luminosities and scales\citep[\eg ][and
references therein]{Bol++08}. Understanding how the stellar and dark matter
``conspire'' to produce this end result is a challenge for galaxy formation
models. In the next decade we should be looking to extend this study to higher
redshifts and lower masses, to go with advances in the simulations of bulge
formation.

With improved weak shear data, it will be possible to gain insights into
misalignments between galaxy halos, and their disks and bulges: this is a key
measurement that tells us about how galaxies are affected by the larger-scale
environment as they form and evolve.   This measurement is difficult with
current data \citep{Man++06b}: having a  larger statistical sample will reduce
shape noise, while surveys designed for weak lensing measurements will allow
for better control of the systematics. Strong lensing observables can also
provide insights on misalignments between halos and galaxy stellar components
on smaller scales.   This information may be particularly interesting in the
case of disk lenses, where the orientation of the stellar component can be
unambiguously determined. Although the vast majority of known strong lenses
are massive ellipticals, several lower-mass disk lenses are known and many
more will be found \citep[\eg][Marshall et al. in prep]{Yor++05,Bol++08}. 

What is needed for these galaxy structure measurements? Current ground-based
weak lensing studies have been limited to massive galaxies at relatively low
redshift.  Deeper or space-based surveys (\eg GEMS, AEGIS) provide higher
background densities but have had limited precision due to the small survey
areas covered. What is needed is a wide-field imaging survey that is deep
enough to extend significant weak lensing measurements of halo structure to
both lower-mass (and more varied type) galaxies and to higher redshifts;
accurate  photometric redshifts will be especially important to cleanly
separate the sources and identify the lenses.

The main obstacle to using strong lenses to probe galaxy-scale dark matter
halos is that these systems are rare: currently there are only $\sim$200 known
galaxy-scale strong lenses.  It would be possible to detect $10^{4-5}$
galaxy-scale lenses with a $\sim10^4$~square degree optical imaging survey of
the kind needed for the high fidelity weak lensing measurements.   The number
of lenses detected is quite a strong function of angular resolution: with
space-based instrumentation (0.1~arcsec PSF FWHM)   we expect some 10 lenses
per square degree \citep{MBS05,Fau++08}, while from the ground we expect to do
perhaps an order of magnitude worse  \citep[][Gavazzi et al. in
prep]{Cab++07}. A $10^4$ square degree  JDEM survey in the optical (where the
source galaxies are bright) to a depth of 27th magnitude would therefore 
discover $10^{5}$ galaxy-scale lenses; LSST would likely see 5 times fewer
than this in a $20000$ square degrees to comparable depth, but would provide
valuable time delays and microlensing signals for all of several thousand
lensed quasars and supernovae. The powerful synergy afforded by the Square
Kilometre Array in the radio is discussed in the white paper by L.~Koopmans.

There is also the prospect of detecting an entire dark galaxy in such a
survey. These are not predicted to occur by the current simulations, at least
not at typical halo masses and luminosities. Finding one -- by focusing on
detecting arc-like features unassociated with foreground light
\citep[\eg][]{S+B07} --  would provide a significant challenge to the world
model. 

High accuracy strong lensing studies not only need large samples to map out
the scatter and probe evolution, but also some intensive follow-up effort.
High spatial resolution IFU spectroscopy will be particularly valuable,
allowing the kinematic structure of the lenses to be mapped out and used as
complementary mass constraints \citep{B+K07,Czo++08}. Separating the stellar
mass components of the galaxy needs high resolution infra-red imaging. We can
imagine making all these observations routinely either with next-generation
adaptive optics on 8-10m class telescopes  from the ground, or with JWST from
space.


\subsection*{Satellites of galaxies}
      
\begin{figure*}[!t]
\centering
\begin{minipage}{0.35\linewidth}
\centerline{\includegraphics[width=\linewidth]{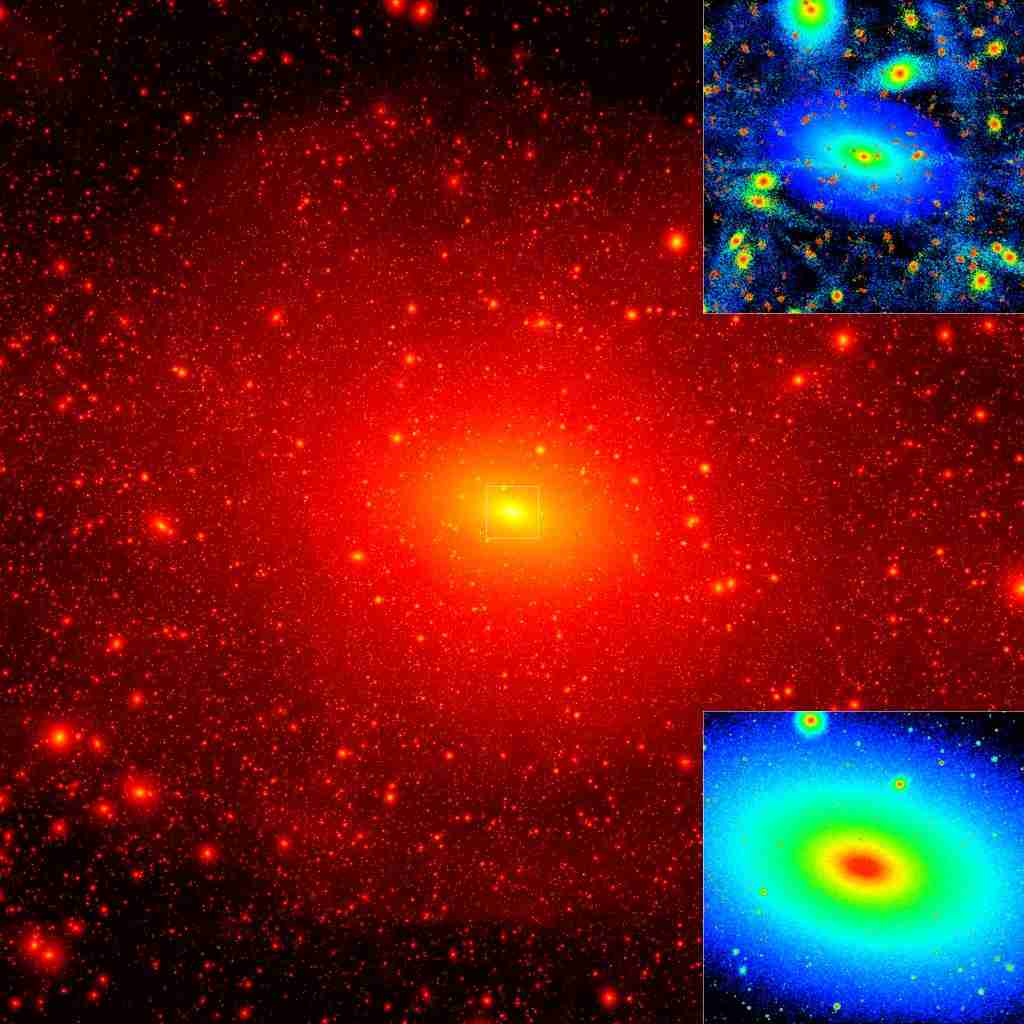}}
\end{minipage}\hfill
\begin{minipage}{0.24\linewidth}
\centerline{\includegraphics[width=\linewidth]{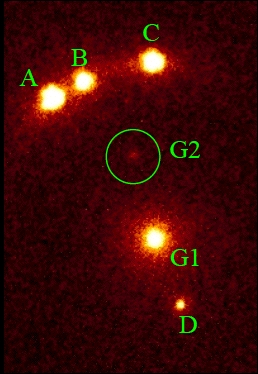}}
\end{minipage}\hfill
\begin{minipage}{0.35\linewidth}
\centerline{\includegraphics[width=\linewidth]{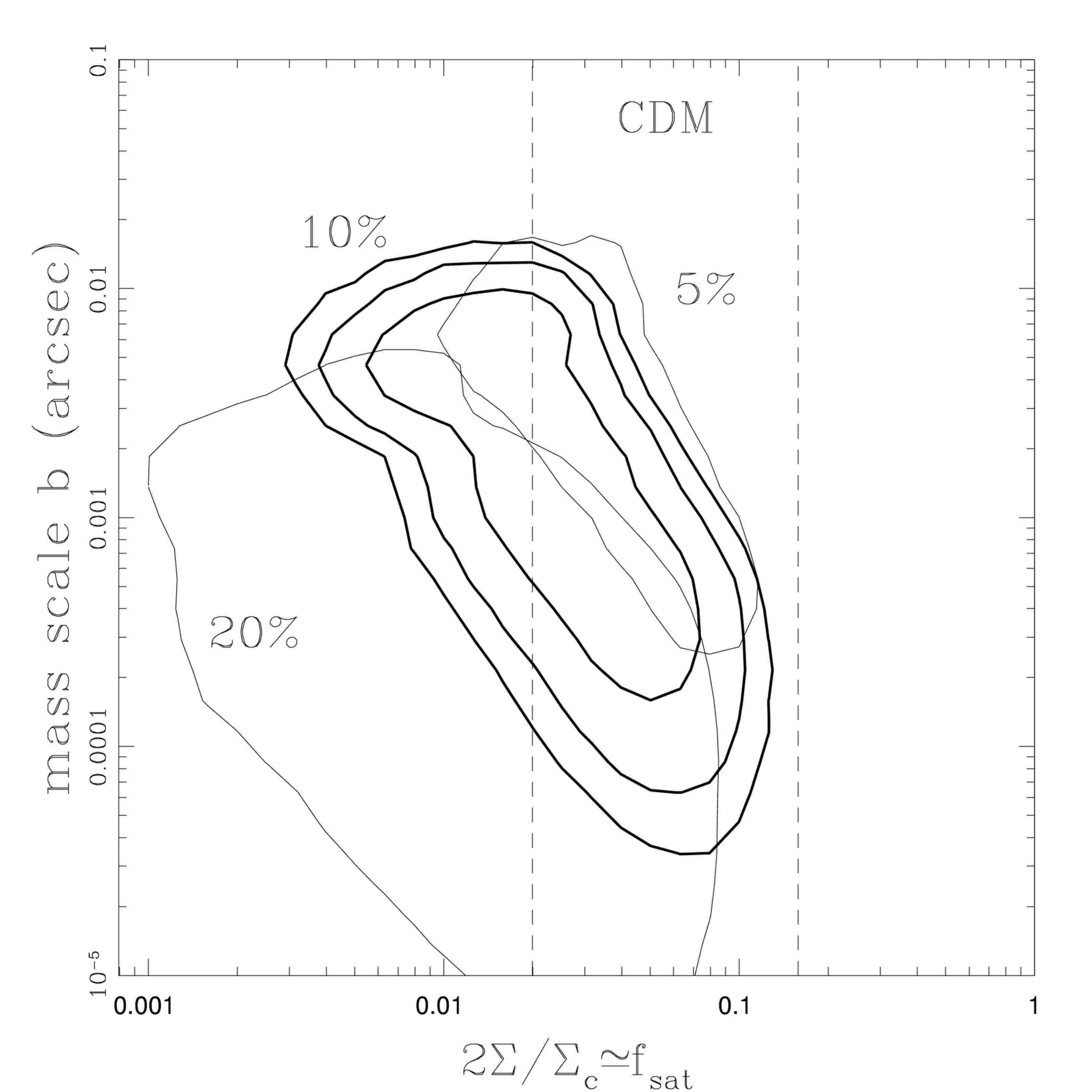}}
\end{minipage}
\caption{Left: high resolution
numerical simulation of a galaxy-scale dark matter halo, showing the 
abundance of small-scale mass structure \citep{Die++08}. Center: 
Keck adaptive optics image of B2045+265 showing the
lensing galaxy (G1) and four lensed images of the
background AGN (A--D).  This is one of the most extreme anomalous
flux ratio systems known: image B should be the brightest of the
three close lensed images, but instead it is the faintest, suggesting
the presence of a small-scale perturbing mass.  The adaptive optics
imaging reveals the presence of a small satellite galaxy (G2) that
may be responsible for causing the anomaly.  Image from \citet{McK++07}.
Right: inferred
constraints on the fraction of galaxy mass in substructure, and the
characteristic sub-halo mass, from an ensemble of 7 lenses showing anomalous
flux ratios \citet{D+K02}.}
\label{fig:satellites}
\end{figure*}

It is on sub-galactic scales that CDM has been tested least rigorously: there
remains the key question of the distribution of mass on sub-galactic scales.
Recently there have been claims of the so-called substructure crisis being
resolved, as the SDSS survey has turned up more and more Milky Way satellite
galaxies: the remaining shortfall in nearby low-mass companion sub-halos is
attributed to star formation inefficiency \citep[see \eg ][]{S+G07,Bus++09}.
This is perhaps a good example of how accepted CDM has become: the problem is
assumed to be in the star formation physics, not the underlying dark matter
model that still predicts 4 times as many dwarf satellites as have been
detected! (Figure~\ref{fig:satellites}.)  Modifications to CDM have been
suggested that predict fewer satellites, either due to a lower mass for the DM
particle (WDM) or an altered interaction cross-section (SIDM). 
Measuring the mass function of galaxy sub-halos is a way of probing the
fundamental properties of dark matter;  future optical imaging surveys more
sensitive than SDSS and covering greater sky area will allow the direct
detection of more faint MW satellites, and to probe their stellar kinematics. 

However, Milky Way studies will very soon become cosmic variance-limited.
There is a clear need for a complementary approach, both to detect dark 
matter substructures independently of their luminous component, and also to
extend the study to galaxies beyond our own. Strong lenses satisfy both these
criteria. Galaxy-scale strong lenses showing almost complete Einstein rings
can be used to search for dwarf satellites via the small perturbations these
structures have on the ring morphology: subhalos with as little mass as
$10^7$~M$_{\odot}$ are expected to be directly detectable in this way with
high signal-to-noise, high resolution imaging data \citep{V+K09}.  The
normalization of the mass function can be constrained by statistical analysis
of the residual images for large numbers of lenses.   High resolution imaging
with the next generation of AO instrumentation with 8-10m class telescopes
would be the tool of choice, exploiting their improved astrometric accuracy
over HST and JWST, and observing in the infra-red to maximize sensitivity to
the old stellar populations of any luminous satellites. As lens samples grow,
30-m class telescopes would enable the follow-up to better keep pace with the
surveys.

Lensed quasars (Figure~\ref{fig:satellites}) offer a similar but perhaps less
expensive route to the subhalo mass function, along with the density profile
and spatial distribution of subhalos. Indeed, this one approach is expanded
upon in  more detail in the white paper by L.~Moustakas. There are three
observables in a lensed quasar that are affected by ``milli-lensing'' by
satellite galaxies: the fluxes, positions, and time delays of the images. 
Flux ratio perturbations are the easiest to measure (provided that
microlensing can be modeled out over many years of cadenced surveying
campaign).  They primarily constrain the fraction of the galaxy mass contained
in substructure, and to some extent the density profile of the subhalos
themselves \citep{D+K02,S+E08}. Astrometric perturbations are smaller,
$\simeq$10 mas \citep{Che++07}, but measurable with AO imaging on 30-m class
telescopes and with VLBI in the radio.  They are useful in combination with
flux ratio perturbations because they measure a different moment of the
subhalo mass function.  Substructure perturbations to time delays are at the
level of hours to days \citep{K+M08}, and may require a dedicated
high-precision photometric lens monitoring mission.  They provide yet a third
moment of the subhalo mass function, and more sensitivity to the spatial
distribution of subhalos around the lens galaxy.  The joint analysis of flux
ratios with image positions and perhaps time delays would test not only
predictions about the subhalo mass function, but also predictions that
subhalos should follow (almost) the same universal density profile and
concentration-mass relation as larger galaxy and cluster halos
\citep{Spr++08}, and that subhalos may be (anti)biased with respect to the
smooth dark matter halo.

Quasar lenses, although rarer than galaxy-galaxy strong lenses by a factor of
10 or more, can be efficiently detected optical imaging surveys by using the
variability of the lensed objects \citep[\eg][]{Koc++06}, allowing
ground-based surveys to recover more small-separation lenses than they would
have otherwise resolved. This places a constraint on any survey we might be
considering: that it be built up from multiple exposures taken over several
years (to give the sources the chance to vary).


\subsection*{Conclusions}

We expect that studies of dark matter structures outside the local group  will
remain one of the major lines of inquiry into this mysterious component of the
universe, complementing direct searches and other astroparticle physics
observations. For measuring the mass distributions of clusters and galaxies,
and their respective subhalo populations, gravitational lensing is the tool of
choice, particularly as  we push out to higher redshifts to probe the
evolution of these structures. 
A wide-field, multi-filter, preferably multi-epoch
optical imaging survey, such as that proposed with LSST or a suitably-designed
JDEM observatory, would enable a wide
range of dark matter science,  based on the (stackable) weak lensing signals
from all halos, and the strong lensing effects due to some. 
To fully exploit these surveys, considerable numbers of supporting
observations will be required, primarily 
including high resolution imaging and high
throughput multi-object spectroscopy. The next generation of ground-based
adaptive optics instrumentation should be capable of providing this support.


\subsection*{References}
\vspace{-2\baselineskip}
\bibliographystyle{SciBook}
\begin{multicols}{2}
\bibliography{dm}
\end{multicols}


\end{document}